\documentclass[aps,prb,showpacs]{revtex4}
\usepackage{amsmath, amssymb,color,graphicx,subfigure}
\usepackage{float}

\newcommand{\beq}{\begin{eqnarray}}
\newcommand{\eeq}{\end{eqnarray}}

\begin{document}

\title{  Modeling   Anisotropic  Plasmon Excitations in Self-Assembled Fullerenes }

\author{Andrii Iurov$^{1}$, Godfrey Gumbs$^{1,2}$, Bo Gao$^{1}$ and Danhong Huang$^{3}$}
\affiliation{$^{1}$Department of Physics and Astronomy,  Hunter College  of the
City University of New York, 695 Park Avenue, New York, NY 10065, USA\\
$^{2}$ Donostia International Physics Center (DIPC),
P de Manuel Lardizabal, 4, 20018 San Sebastian, Basque Country, Spain\\
$^{3}$Air Force Research Laboratory, Space Vehicles Directorate,
Kirtland Air Force Base, NM 87117, USA}

\date{\today}

\begin{abstract}
The plasmon excitations in Coulomb-coupled spherical two-dimensional electron gases (S2DEGs)
reveal an interesting  dependence on the displacement vector between the centers of
the spheres with respect to the axis of quantization for the angular momentum quantum number $L$.
Specifically, plasmon modes for a bundle of three
S2DEGs have been obtained within the random-phase approximation (RPA). The inter-sphere
Coulomb interaction matrix elements and their symmetry properties were
also investigated in detail. The case of a bundle gives an adequate picture
of the way in which the Coulomb interaction depends on the orbital angular momentum quantum number
$L$ and its projection $M$. We concluded that the interaction  between the S2DEGs aligned
at an angle of $45^{\rm o}$ with the axis of quantization is negligible compared
to the interaction along and perpendicular to the quantization axis, which are
themselves unequal to each other. Consequently,
the plasmon excitation frequencies reveal an interesting orientational anisotropic
coupling to an external electromagnetic field probing the charge density oscillations.
Our result on the spatial correlation may
 be experimentally observable. In this connection,  there have already been
some experimental reports pointing to a similar effect in nanoparticles. 

\end{abstract}


\maketitle

\section{Introduction}
\label{sec1}

The year 1995 was the culmination of several years of the search for    fullerene materials.\,\cite{O1,O1b,O2,O3,O4,O5,O6}
It also marks the beginning of a whole new area of investigation into the optical and thermal
properties of fullerene aggregates. From both an experimental and theoretical
perspective, this branch of condensed matter physics is receiving international
attention. Consequently, the theory for the properties of interacting
fullerenes is a subject of considerable interest.\,\cite{1,2,3}
\medskip

\begin{figure}[ht!]
\centering
\includegraphics[width=0.3\textwidth]{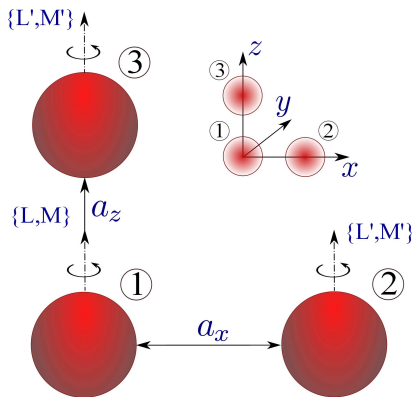}
\caption{(Color online) Schematic illustration of
a triad of displaced S2DEGs. The axis of quantization is
along the $z$ direction with angular momentum quantum
number $L$ and component $M$. One shell
has its center at the origin and the others are centered on
the $x$ and $z$ axes.}
\label{FIG:1}
\end{figure}

Following  the recent advances in production techniques such as solvent-assisted
self-assembly,\cite{XXX2} fullerenes can now be produced in specific quantities
even to form thin films of fullerene-like MoS     nanoparticles\cite{XXX3}
and those achievements have stimulated  renewed interest in these materials.
Additionally, the ability to control optical fields has now made
it feasible to  ascertain the plasmon excitations in pre-arranged
arrays of fullerenes.\cite{XXX2,XXX3}
The present paper has been stimulated by these exciting new developments, even though
it employs a relatively simple model which is reasonable when the separation between
the energy bands is large.\,\cite{Apell} Additionally, our results may serve as a guide for
explaining experimental studies.\,\cite{expt1,expt2} Specifically, we are interested in calculating
the plasma excitations of Coulomb coupled spherical two-dimensional electron gases
(S2DEGs).\,\cite{Lett1,Lett2,SSC,SSC5} Here, we model three coupled spherical shells
as shown in Fig.\,\ref{FIG:1}.
In our discussion below, we refer to the S2DEG at the origin as ``$1$"', and  the
one centered  on the $x$ and $z$-axis as ``$2$'' and ``$3$'', respectively.
This notation then allows us to label  the Coulomb matrix elements between  spherical
shells ``$i$'' and ``$j$'' with subscript ``$i-j$'', where $i,j=1,\,2,\,3$.
\medskip

 Electron energy loss spectroscopy (EELS) has been used to probe
the plasmon excitations for concentric-shell fullerenes
embedded  in a film.\cite{EELS} Furthermore,  perfectly spherical
 shells were used in the theoretical modeling of the EELS data
and the agreement was good.  The model of Lucas, et al.\cite{Lucas}
was shown to be qualitatively adequate for understanding the optical
data for multi-shell fullerenes. In that work,\cite{Lucas}
 the ultraviolet dielectric tensor of monolayer graphene is adapted
 to  the
spherical geometry of a fullerene by averaging over the three
possible orientations of the ${\bf c}$ axis.  Thus, a continuum
model was used by Lucas, et al.\cite{Lucas} starting from the
planar local di electric function of planar monolayer graphene.

\medskip

Each S2DEG may be polarized by external electromagnetic fields. However, the S2DEG is only
polarized for finite angular momentum quantum number $L\neq 0$. Although $L$ is a good
quantum number for labeling the plasma excitations on an isolated shell, it is not the
case for displaced, coupled  S2DEGs. In contrast to the well-understood situation of determining
the plasma excitations in fullerenes,\,\cite{Apell} we seek its properties
when the rotational symmetry is broken. Here, for simplicity, $L=0$ corresponds
to a non-circularly-polarized probe field, while $L=1$ is associated with a
circularly-polarized probe field. Moreover, the higher angular momentum with $L>1$
may be achieved by a special light beam, e.g., a helical light beam.
However, when two S2DEGs have their centers well separated so
that there is no overlap of their charge distributions, the breaking of
the spherical symmetry leads to significant differences with the
multi-shell buckyball. Similar calculations of the polarization functions proved
the existence of strongly localized image states near the surface of a buckyball \cite{ours}. Anisotropy of the plasmon interaction could be seen for two interacting shells \cite{ours2}
\medskip

The topology of multiple shells forming an aggregate determines
the properties of Coulomb interaction and consequently the details
of the collective plasma excitations. For a single S2DEG, the Coulomb
interaction depends on the angular momentum quantum number $L$ which in
itself together with the radius of the shell serves to define the frequency of
the plasmon mode. In contrast, when a pair of S2DEGs are coupled for a configuration
such as the one  exhibited in Fig.\,\ref{FIG:1}, the polarization functions
for all values of $L$ on each sphere may not be dissociated from one another.
The inter-sphere Coulomb matrix element
depends on both $L$ and its projection $M$ on the axis of quantization.
Therefore, in principle, the plasmon mode equation is given in terms of a
determinant of infinite order.  But, for three interacting S2DEGs, the associated
matrix may be broken down into diagonal sub-matrices corresponding to
$L=1,\,2,\,3,\,\cdots$ and consisting of $3(2L+1)\times 3(2L+1)$ elements
whose Coulomb interactions depend on $L$ and $(2L+1)$ values
of $M$ for each of the three shells.
In contrast, the off-diagonal sub-matrices
depend on Coulomb matrix elements which are functions of pairs of different
angular momenta, $L$ and $L^\prime$, arising from each sphere. These off-diagonal
Coulomb terms are generally smaller than their diagonal
counterparts and so may be formally treated as perturbations. Consequently, in
the lowest-order approximation, the $L=1$ mode is split
by Coulomb interactions depending on $M=0,\,\pm 1$  from each
sphere, leading to the occurrence of nine plasmon branches.
However, some of these may be very close to each other depending on the strength
of the Coulomb interaction. For example, when two S2DEGs are aligned along
the $x$-direction, the number of well-separated plasmon branches is three and for the $z$-alignment we
found only two non-equivalent plasmon branches.
\medskip

In Sec.\,\ref{sec2}, we will first formulate the method for
calculating the plasmon equations on three coupled S2DEGs.
This is based on the random-phase approximation (RPA) in
evaluating the induced density fluctuations for a weak external
perturbation. Section\ \ref{sec3} is devoted to a discussion
of our numerical results. Some concluding remarks are given in
Sec.\,\ref{sec4}.

\section{General Formulation of the Problem}
\label{sec2}

Let us consider a bundle consisting of three spherical shells
with their centers located at the origin, on the $x$ axis and
the $z$ axis. That is, the center of one of the spheres is
at $\textbf{r}_1=0$ with radius $R_1$ whereas the others   are centered
at $\textbf{r}_2=(a,0,0)$, with radius $R_2$ and
$\textbf{r}_3=(0,0,a)$,  with radius $R_3$. We assume that the
inequalities $a>R_1+R_2$ and $a>R_1+R_3$ are satisfied. In the
absence of tunneling between the shells, the wave function
for an electron on the $j$-th shell ($j=1,2,3$) is given by

\begin{equation}
<{\bf r}\mid j\nu>=\Psi_{jlm}\left(\vec{r}-(3-j)(j-1)a\hat{e}_x
-\frac{1}{2}(j-1)(j-2)a\hat{e}_y\right),
\hspace{.3cm}
\Psi_{jlm}(\vec{r})=f_j(r)\frac{1}{\sqrt{ R_j^2}}
 Y_{lm}(\Omega)    \ ,
\label{tmz.1}
\end{equation}
in terms of spherical harmonics $ Y_{lm}(\Omega)$
with $\nu=\{l,m\}$ and $f_j^2(r)=\delta(r-R_j)$. The
energy spectrum has the form of
$\epsilon_{j,\nu}= \hbar^2l(l+1)/(2m^{\ast}R_j^2)$.
The induced potential
$\Phi$ satisfies Poisson's equation
$\nabla^2\Phi({\bf r},\omega)= (4\pi e/\varepsilon_s)
\delta n({\bf r}, \omega)$,
where $\varepsilon_s\equiv 4\pi\epsilon_0\varepsilon_b$
and $\varepsilon_b$ is the uniform background dielectric
constant. We use linear response theory to calculate the
induced particle density $\delta n({\bf r},\omega)$. After a
straightforward calculation, we obtain

\begin{equation}
\sum_{j^\prime=1}^3\sum_{L^\prime=0}^{\infty}
\sum_{M^\prime=-L^\prime}^{L^\prime}
\left[\delta_{jj^\prime}\delta_{LL^\prime}\delta_{MM^\prime}+
\frac{2e^2}{\varepsilon_s}\Pi_{j^\prime,L^\prime }(\omega)
V_{j^\prime L^\prime M^\prime,jLM}(R_j,R_{j^\prime},a)\right]
U_{j^\prime,L^\prime M^\prime}=0 \ ,
\label{tmz:13}
\end{equation}
where

\begin{equation}
\Pi_{L}(\omega)=  \sum_{l,l^\prime}
\frac{f_0(\epsilon_{l})-f_0(\epsilon_{l^\prime})}
{\hbar\omega+\epsilon_{l^\prime}-\epsilon_{l}}
(2l+1)(2l^\prime+1)
\left( \begin{matrix}
l&l^\prime& L\cr
0 & 0 & 0\cr
\end{matrix}\right)^2\ ,
\label{gae13}
\end{equation}

\begin{eqnarray}
V_{j^\prime L^\prime M^\prime,jLM}(R_j,R_{j^\prime};a)
&=&8\int
\frac{d^3\textbf{q}}{q^2}\
j_L(qR_j)j_{L^\prime}(q R_{j^\prime})
Y_{LM}^\ast(\hat{\textbf{q}}) Y_{L^\prime M^\prime}(\hat{\textbf{q}})
\nonumber\\
&\times& e^{ i[(3-j)(j-1)q_xa  +
\frac{1}{2}(j-1)(j-2) q_ya ]  }e^{- i[(3-j^\prime)(j^\prime-1)q_xa  +
\frac{1}{2}(j^\prime-1)(j^\prime-2) q_ya ]  }
\label{tmz:14}
\end{eqnarray}
with $j_L(x)$ being a spherical Bessel function and
$\Pi_{j,L}(\omega)$ being the density response function of
the $j$-th nano shell. Also, we have introduced the quantity

\begin{equation}
U_{j,LM} = \frac{1}{L_xL_yL_z}\sum_{q_x,q_y,q_z}e^{ i[(3-j)(j-1)q_xa  +
\frac{1}{2}(j-1)(j-2) q_ya ]  }
\frac{\delta n(q_x,q_y,q_z,\omega)}{q_x^2+q_y^2+q_z^2}
\ j_L\left(q R_j\right)
Y_{LM}^\ast(\hat{\textbf{q}}) \ .
\label{tmz:12}
\end{equation}
Here, $L_x,\,L_y$ and $L_z$ are normalization lengths. Thus, non-trivial solutions for
the charge density oscillations correspond to the zeros of the determinant
of the coefficient matrix for $U_{j,LM} $ in Eq.\,(\ref{tmz:13}).

\section{Numerical Results}
\label{sec3}

\begin{figure}[ht!]
\centering
\includegraphics[width=0.35\textwidth]{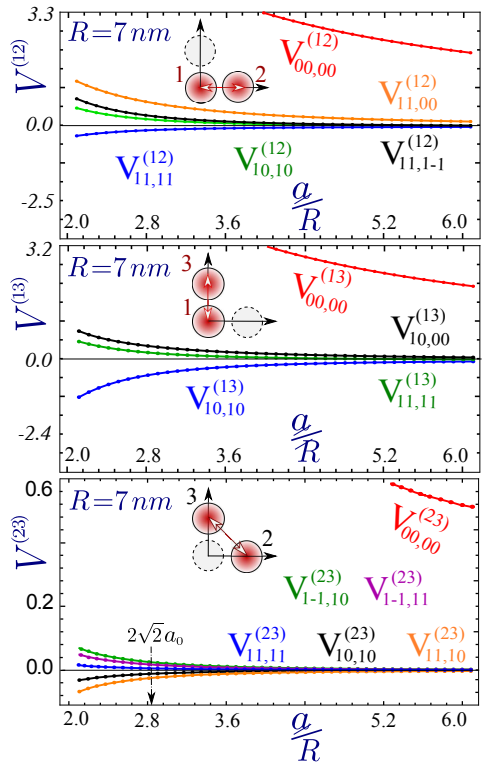}
\caption{(Color online) Coulomb matrix elements, given in units of
$2e^2/(\pi\epsilon_s R)=0.59$\,eV,  for the various interactions between spheres
(a) ``$1$'' and ``$2$'', (b) ``$1$'' and ``$3$'' and (c) ``$2$'' and ``$3$''.
The radii of the S2DEGs are  equal with $R=7$\,nm. The inter-sphere distance $a$
is given in units of $R$.}
\label{FIG:3}
\end{figure}

In Fig.\,\ref{FIG:3}, we compare the strengths of the electrostatic interaction
between pairs of S2DEGs when $L=0,M=0$ and $L=1$ with $M=0,\pm 1$, where all
the non-zero Coulomb matrix elements for an interacting triad of spheres
displayed schematically in Fig.\,\ref{FIG:1} are presented.
The matrix elements $V_{LM,L^\prime M^\prime}$ with $L=0$ and $L^\prime=0$, which
do not contribute to plasmon excitation, are also provided for comparison.
These matrix elements greatly exceed all the others because of rotational symmetry.

\medskip

Our calculations show that the Coulomb interaction is significant only
at distances that are comparable  to the sum of the radii of the two interacting spheres.
In the notation of Fig.\,\ref{FIG:1}, we have shown that the $(2-3)$ interaction
is negligible compared to the $(1-2)$ and $(1-3)$ interactions. The reason for this difference is
not due to the larger separation between $(2-3)$ which is $\sqrt{2}$ times that between $(1-2)$
or $(1-3)$ but it is a consequence of the angular dependence inherent in the matrix elements.
As a matter of fact, unlike the $(1-2)$ and $(1-3)$  matrix elements, we are unable to obtain
semi-analytic expressions for the $(2-3)$ Coulomb matrix elements since even the angular
part of the related Coulomb integral cannot be expressed in analytical form. Using numerical
methods, however, we have estimated these matrix elements to be on the order of $10^{-2}$ times
smaller than all the rest. This results in the existence of one strong nearly-degenerate plasmon mode,
which is practically independent of separation $a$ between the centers of the spheres.
Additionally, due to the reduced symmetry, there are more non-zero matrix elements
for $(2-3)$ coupling compared to the  $(1-2)$ and $(1-3)$ cases, which however does not improve
the impact of $(2-3)$ interaction because of their relatively small magnitude.
\medskip

Setting $L,\,L^\prime=1$ and $M,\,M^\prime=0,\pm 1$
in the set of linear equations (\ref{tmz:13}) for $j=1,\,2,\,3$, leads
to a $9\times 9$ matrix of the coefficient matrix of $\{ U_{j,LM}\}$.
The dephasing constant $\delta$ is chosen to be $0.05$\,eV$/\hbar$.
We present density plots in the following figures for the inverse of this
dielectric function to demonstrate the plasmon modes of interacting S2DEGs.
\medskip

\begin{figure}
\centering
\includegraphics[width=0.65\textwidth]{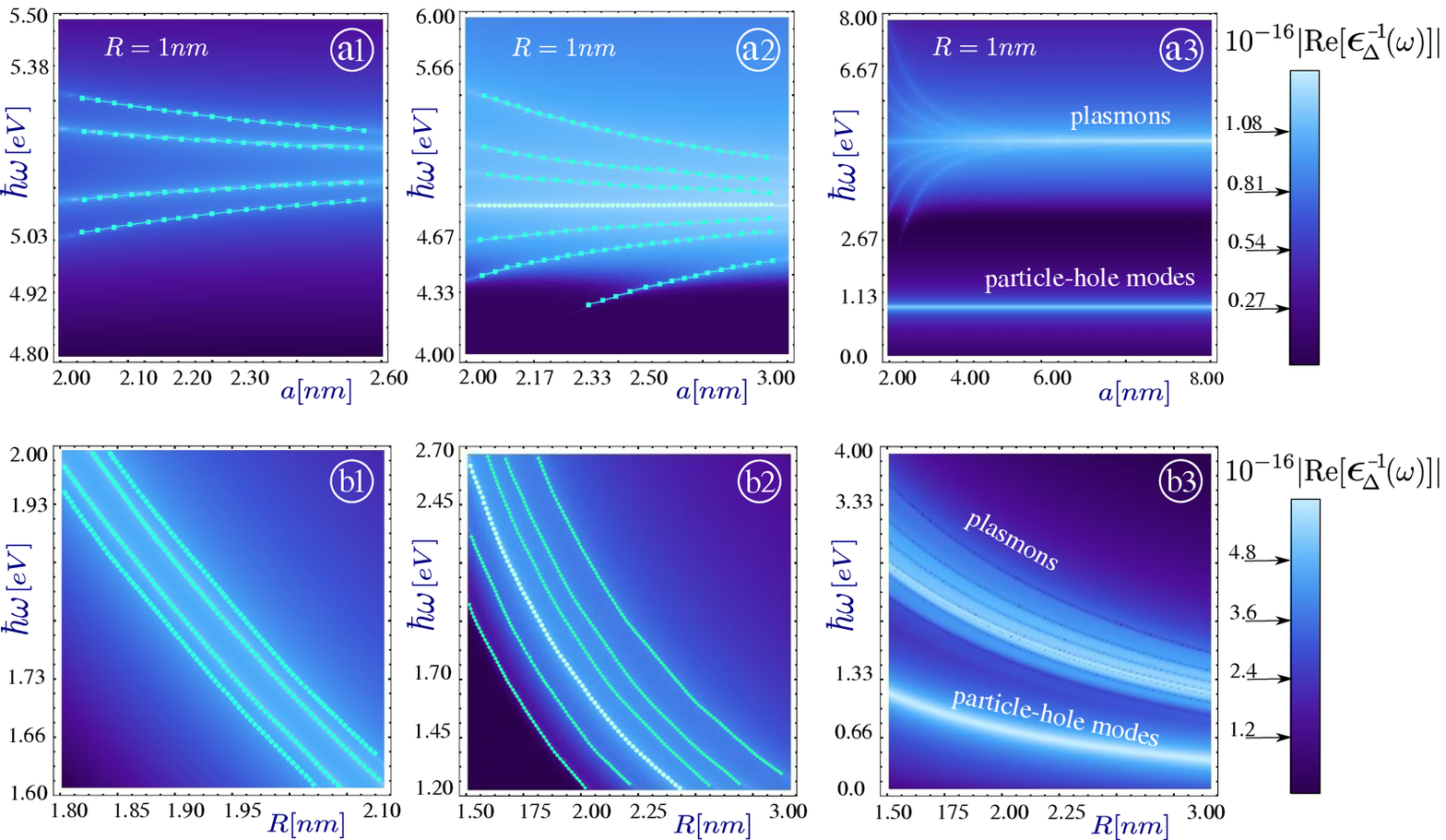}
\caption{(Color online) Density plot of the inverse of the dielectric function
$ \epsilon_\Delta (R,\omega)  $,
defined to be the determinant of the coefficient matrix in Eq.\,(\ref{tmz:13}),
demonstrating the plasmon  modes of interacting spheres. The upper row [plots (a1)-(a3)] shows
how the frequencies of the excitations depend on the separation $a$ between the centers of
the shells with equal radius $R=1$\,nm. These three panels highlight the plasmon spectrum for
different ranges of the variable $a$. The lower panel presents the plasmon frequency dependence
on the radius of each S2DEG.  Plots (a1) and (b1) show the \textit{four} plasmon branches
for the \textit{two} spheres $(1-3)$, on the $z-$axis, for which the matrix could be diagonalized
exactly. Plots (a2) and (b2) show the plasmon branches for three interacting S2DEGs,
including the horizontal nearly-degenerate subset of branches, resulting from $(2-3)$
interaction. All the above mentioned plots also include numerical solutions for the plasmons,
corresponding to the  zeros of the dielectric function. Panels (a3) and (b3) present all the
plasmon branches of an interacting S2DEGs over a wider range of  separation and
frequencies. The number of occupied energy levels used in these calculations is $N_F=8$
for all  six panels.}
\label{FIG:4}
\end{figure}

Now, let us consider three identical S2DEGs with equal radius $R_1=R_2=R_3$
at the corners of a right-angle triangle as shown in Fig.\,\ref{FIG:1}.   Numerical results
for the plasmon and particle-hole frequencies are presented in Fig.\,\ref{FIG:4}, where we  only
chose $L,\,L^\prime=1$ and $M,\,M^\prime=0,\pm 1$. Generally, the number of plasmon
branches is determined by the dimensionality of the truncated matrix, with $2L+1$ solutions
associated with  each sphere. We separate the case of two spheres with $z-$alignment $(1-3)$,
for which the interaction sub-matrix is diagonal with two unequal non-zero elements,
and the whole matrix allows exact diagonalization. Such interaction results in two
\textit{  symmetrically split pairs} of plasmon branches, presented in Fig.\,\ref{FIG:4}\, (a1)
and (b1). All  other plots in Fig.\,\ref{FIG:4} show the plasmons for three interacting
S2DEGs. We expect  \textit{nine} separate plasmon branches, corresponding to the solutions
of $\Re e[\epsilon_\Delta](R,\omega)]=0$. However, \textit{three} of them, corresponding
to the $(2-3)$ interaction, are nearly-degenerate, due to its low relative strength
of the Coulomb coupling. Apart from that, one or a few solutions, located far away from
the central branch, are very weak and could not be shown for the chosen range of frequencies.
Plots $(a3)$ and $(b3)$ in Fig.\,\ref{FIG:4} present both plasmons and particle-hole modes,
as the peaks of $1/\Re e[\epsilon_\Delta(\omega,R)]$ and $\Im m[\Pi^{0}(\omega,R)]$ correspondingly.
We calculated those quantities over a wider range of frequencies for different values of radius
$R$ and separation $a$. We conclude that the plasmons and single-particle excitation regions
are well-separated so that the plasmons are not \textit{Landau damped} for all cases of an interacting
bundle. But, the electron-hole mode  region and plasmons get closer with increasing radius of the shell.
\medskip

\begin{figure}[ht!]
\centering
\includegraphics[width=0.9\textwidth]{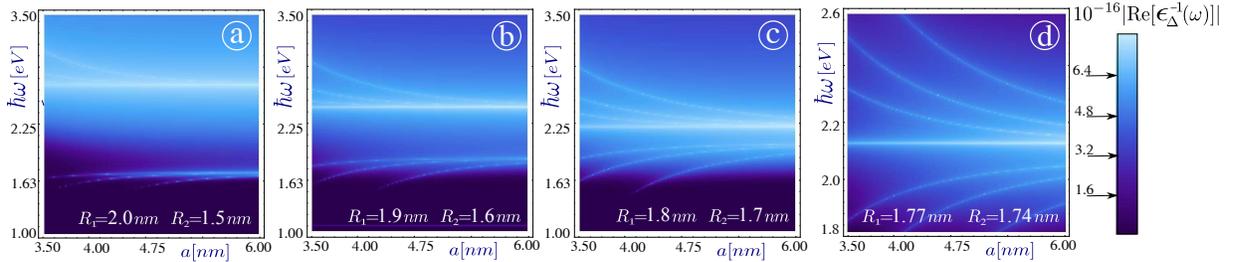}
\caption{(Color online) Density plot of the inverse dielectric function {\em vs.}
the separation between their centers. Each plot corresponds to a separate case of two unequal
radii of the sphere located at the origin $(1)$ and two others with $R_1  \neq R_2 = R_3$.
The plasmon branches correspond to the maxima of $1/\Re e[\epsilon_\Delta]$. We
chose the density such that the number of occupied levels is $N_F=9$ for all spheres. Also,
$L=1$ as the only value of the angular momentum, contributing to plasmon excitations.
All plasmon excitations merge to form a single branch for $a\gg R$.}
\label{FIG:5}
\end{figure}

Our results for plasmon excitations for three interacting S2DEGs with unequal radii are
presented in Fig.\,\ref{FIG:5}. We chose  the radius of the sphere at the origin (sphere $1$) to be
larger  than that  of either sphere $2$ or $3$  whose radius is still chosen to be equal to one another.
i.e., $R_1 > R_2 =R_3$. We note that in this case the ratio of the   Coulomb interaction
between spheres  $(2-3)$ to that between either $(1-2)$ or $(1-3)$   is even smaller, compared to the
same ratio of three S2DEGs with the same radius $R_1=R_1=R_3$  since the distance between
shells $(1-2)$ and $(1-3)$ is increased.
\medskip

When the separation $a$ between spheres $(1-2)$ or $(1-3)$ becomes large compared to $R_1$, the
plasmon frequencies when plotted as a function of $a$ form two horizontal branches,
with each one arising from the plasmons on each isolated sphere  with radii $R_1$ and
$R_2$. The higher frequency branch is due to the two spheres with smaller radius
$R_2=R_3$. As a consequence, this plasmon branch exhibits a stronger resonance  peak with
a higher degree of degeneracy. The three solutions, corresponding to the $(2-3)$ interaction,
stay nearly degenerate for all values of the separation $a$. This behavior
is similar to the previous case when $R_1=R_2=R_3$.
The resulting plasmon branches split apart, creating  three asymmetric
pairs of solutions. This is a general feature.

\begin{figure}[ht!]
\centering
\includegraphics[width=0.3\textwidth]{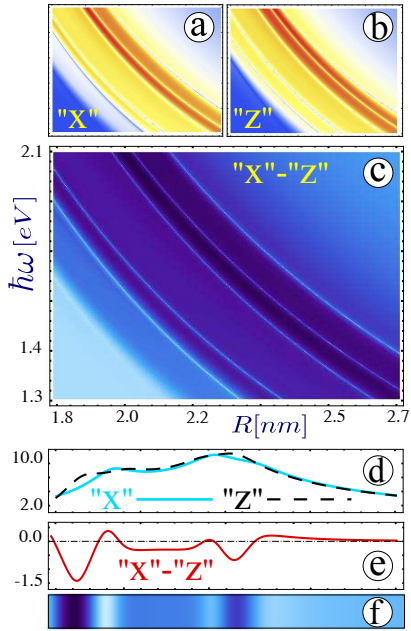}
\caption{(Color online) Comparison of plasmon excitations  for the $(1-2)$ and $(1-3)$ pairs
of S2DEGs. Panels (a) and (b) are density plots for the inverse dielectric function
for the $(1-2)$ and $(1-3)$ interacting S2DEGs, respectively. Plot (c) shows the difference
between the absolute values obtained in (a) and (b) over the same range of  radius
($R_1=R_2=R$) and plasmon frequencies ($\omega_{\rm min}=1.3$\,eV$/\hbar$ and
$\omega_{\rm max}=2.1$\,eV$/\hbar$).
Panel (d) shows the $R-$dependence of the inverse
dielectric function for chosen frequency $\omega=1.50$\,eV$/\hbar$, which is outside the
frequency range of panel (c). Panel (e) demonstrate the difference between these
results in (d). The results presented in the form of $\ln(1+\Pi)$ in order to minimize
the difference between the lowest and the highest values. Panel (f) represents
the density plot, for the preceding plot (e), which also could be considered as
the colorbar for panel (c).}
\label{FIG:1+}
\end{figure}

\medskip
Figure\ \ref{FIG:1+} compares  the plasmon excitations  for the $(1-2)$ and $(1-3)$ pairs
of S2DEGs.  From the results presented here, we conclude that we have
clearly demonstrated the anisotropy of the plasmon mode excitations  for a particulate.
The sum-rule presented by Apell, Echenique, and Ritchie \cite{AER} for  surface
plasmon frequencies does not apply here.   Although the surfaces in our model
 exhibit spatially localized collective modes,  the plasmons for an isolated S2DEG may be
 labeled by a single quantum number $L$ but, when the S2DEGs are coupled, all angular
  momentum quantum numbers and their  projections onto the axis of quantization
determine the plasmon frequencies.
However, we    emphasize that these plasma frequencies  for the $(1-2)$ and $(1-3)$
 configurations are functionals of the  plasmon frequency $\Omega_L$
 of a single S2DEG, and consequently must  be related to each other.
 The frequency    $\Omega_L$ is determined by the solutions of the
equation
 $\Re e  \left[1-e^2/(\epsilon_s (2 L+1)R) \Pi_L(\Omega_L)\right]=0$.
Due to the diagonal nature of the Coulomb  interaction submatrices for the
$(1-3)$ case of $z$-alignment, the equations for the plasmon frequencies may be reduced to
$\text{Re}\left[1-e^2/(\epsilon_s (2 L+1)R) \Pi_L(\omega_z) \pm v_{1,3}(a) \Pi_L(\Omega_z)\right]=0$,
leading to the \textit{two pairs} of solutions which  obviously depend on $\Omega_L$.
For the case of $(1-2)$ $x$-alignment, the Coulomb potential matrix elements contain a larger
number of non-zero teems and, more importantly, this submatrix is not diagonal.
Consequently,    the plasmon  equation may not be presented in such a simple functional
form. Complexity of the resulting plasma equation does not  prevent numerical solution
but clearly highlights the difference in the plasmon frequencies for $x$ and $z$-alignments.

\section{Concluding Remarks}
\label{sec4}

In this paper, we presented a theory for calculating the orientational dependence
of coupled plasmon excitations on a pair of interacting S2DEGs.  Though the model
is simple, we demonstrated that it  is a powerful  tool for exhibiting the novel
orientational dependence of plasmon excitations.
The energy band structure is
a suitable approximation for fullerenes when the energy bands are far apart. However,
realistic energy band structure may be incorporated into the polarization function
through a  form factor involving the wave functions of the electrons as well as
the Coulomb matrix  elements.
\medskip

We note that  spectral correlations  have been observed experimentally for metallic nanoparticles
\,\cite{Yang1} In that work,  the plasmons for pairs  were studied using polarization-selective
total internal reflection.  Their measurements show that the frequencies for the coupled plasmon
modes  depend on whether the incident light wave vector perpendicular and parallel to the dimer
axis. Related work on dimer plasmons has been conducted by Nordlander, et al. (Ref.[\onlinecite{Nord1}])
with the conclusion  that the hybridized plasmon energy arising from individual metallic nanoparticles
is determined by the  orientation of the inter-particle axis with respect to the axis of polarization
of the two constituents modeled as incompressible spherical liquids. Although  our model differs from that
in Ref.[\onlinecite{Nord1,Prodan}] the conclusions about the existence of anisotropy in the plasmon excitation
energies in these systems are in agreement. Similar effects are also expected to be observed in the case of
nano-eggs: non-concentric multishells of nanoparticles. The hybridization of the plasmons has been proven to
be an adequate and precise method to describe the plasmonic structure \cite{Wu,Bardhan}. The field enhancement,
corresponding to the resonant excitation of plasmons, was reported to be much larger in the case of concentric
nanoparticles, which support indirectly the concept of plasmon spatial correlation.
\medskip

Since the plasmon dependence on the relative orientation and the  separation
between the interacting spheres was the principal property under investigation,
we considered an aggregate of three which includes coupling along,
perpendicular and making an angle with the axis of quantization for angular momentum.
We clearly established the orientational dependence of the coupled plasmon excitation
frequencies arising from the sensitive nature of the Coulomb matrix elements.
Our formalism allows for any number and location of the S2DEGs, thereby leading to
possible studies of aggregates of different size and shape.
\medskip

Once the angular momentum quantum number $L$ is specified, the number of plasmon branches
is determined by all the different non-zero potential matrix elements. The number  of those
elements is a consequence of  symmetry  properties.  Specifically, when a pair of  spheres
($(1-3)$) are aligned along the axis of quantization, there are only two such non-zero
matrix elements, whereas three of them exist for $(1-2)$ interaction between the spheres,
aligned along the  $x-$axis. There are five non-zero Coulomb matrix elements between spheres
whose centers lie on a line making an angle of forty-five degrees with the axis of quantization,
but much smaller in magnitude. Additionally,  the effect of the Coulomb interaction on the
plasmons of a smaller sphere,  occurring  at a higher frequency,  is just a small perturbation
for  $(2-3)$ coupling. Our conclusion is that the unique characteristic has high potential for
electronics, sensing applications and photoelectric conversion.

\begin{acknowledgments}
This research was supported by  contract \# FA 9453-07-C-0207 of
AFRL. DH would like to thank the support from the Air Force Office of Scientific Research (AFOSR).
\end{acknowledgments}


\begin{thebibliography}{00}


\bibitem{O1}E. Osawa, Kagaku (Kyoto) {\bf 25}, 854 (1970) [in Japanese].

\bibitem{O1b}J. F. Anacleto and M. A. Quilliam,
Anal. Chem. {\bf 65}, 2236 (1993).

\bibitem{O2}H. W. Kroto, J. R. Heath, S. C. O'Brien, R. F. Curl and R. E. Smalley,
Nat. {\bf 318}, 162 (1985).

\bibitem{O3}S. Iijima,
J. Crystal Growth {\bf 50}, 675 (1980).


\bibitem{O4}P. R. Buseck, S. J. Tsipursky and R. Hettich,
Sci. {\bf 257}, 215 (1992).

\bibitem{O5}J. Cami, J. Bernard-Salas, E. Peeters and S. E. Malek,
Sci. {\bf 329}, 180 (2010).

\bibitem{O6}C. A. Poland, R. Duffin, I. Kinloch, A. Maynard, W. A. H. Wallace, A. Seaton, V. Stone, S. Brown, W. MacNee and K. Donaldson,
Nat. Nanotechn. {\bf 3}, 423 (2008).

\bibitem{1}T. Inaoka,
Surf. Sci. {\bf 273}, 191 (1992).

\bibitem{2}J. Tempere,  I. F. Silvera and J. T. Devreese,
Phys. Rev. B {\bf 65}, 195418 (2002).

\bibitem{3}C. Yannouleas, E. N. Bogachek and U. Landman,
Phys. Rev. B {\bf 53}, 10225 (1996).

\bibitem{XXX2}   Lang Wei, Jiannian Yao, and Hongbing Fu,
ACS Nano, 7 (9), pp 75737582 (2013).

\bibitem{XXX3}  Manish Chhowalla and Gehan A. J. Amaratunga,
Nature \textbf{407}, 164  (2000).

\bibitem{Apell}D. \"Ostling, P. Apell and A. Rosen,
Europhys. Lett. {\bf 21}, 539 (1993).

\bibitem{expt1}G. Gensterblum, J. J. Pireaux,  R. Gaudano, J. P. Vigneron,
A. A. Lucas and W. Kr\"atschmer,
Phys. Rev. Lett. {\bf 67}, 2171 (1991).

\bibitem{expt2}E. Schmen, J. Fink and W. Kr\"atschmer,
Europhys. Lett. {\bf 17}, 51 (1992).

\bibitem{Lett1}V. K. Voora, L. S. Cederbaum and K. D. Jordan,
J. Phys. Chem. Lett. {\bf 4}, 6 (2013).

\bibitem{Lett2}A. K. Belyaev, A. S. Tiukanov, A. I. Toropkin, V. K. Ivanov,
R. G. Polozkov and A. V. Solov\'yov,
Physica Scripta {\bf 80}, 048121 (2009).

\bibitem{SSC}P. Longe,
Solid State Commun. {\bf 97}, 857 (1996).

\bibitem{SSC5}M. T. Michalewicz and M. P. Das,
Solid State Commun. {\bf 84}, 1121 (1992).

\bibitem{EELS}
L. Henrard,* F. Malengreau,  P. Rudolf, K. Hevesi,
R. Caudano, and Ph. Lambin,  Th. Cabioch,
\prb {\bf 59},  5832 (1999).

\bibitem{ours} Godfrey Gumbs, Antonios Balassis, Andrii Iurov, and Paula Fekete,
 The Scientific World Journal, vol. {\bf 2014}, 726303 (2014).

\bibitem{ours2} Godfrey Gumbs, Andrii Iurov, Antonios Balassis, Danhong Huang; arXiv:1309.7328v3 [cond-mat.mes-hall].

\bibitem{Lucas}   A. A. Lucas, L. Henrard, and Ph. Lambin, Phys. Rev. B 49, 2888
 (1994).

\bibitem{AER}  S. P. Apell, P. M. Echenique, and R. H. Ritchie,
Ultramicroscopy {\bf 65}, 53 (1996).

\bibitem{Yang1} Yang, Shu-Chun and Kobori, Hiromu and He, Chieh-Lun and Lin, Meng-Hsien and Chen, Hung-Ying and Li,
Cuncheng and Kanehara, Masayuki and Teranishi, Toshiharu and Gwo, Shangjr, Nano Letters, {\bf 10}, 2 (2010).

\bibitem{Nord1} Nordlander, P. and Oubre, C. and Prodan, E. and Li, K. and Stockman, M. I., Nano Letters, {\bf 4}, 5 (2004).

\bibitem{Prodan}  Prodan, E. and Radloff, C. and Halas, N. J. and Nordlander, P., Science, {\bf 302}, 5644 (2003).

\bibitem{Wu} Y.Wu and P.Nordlander, The Journal of Chemical Physics, {\bf 125}, 125, 124708 (2006).

\bibitem{Bardhan} Bardhan, Rizia and Mukherjee, Shaunak and Mirin, Nikolay A. and Levit, Stephen D. and Nordlander, Peter and Halas, Naomi J.,
The Journal of Physical Chemistry C, { \bf 114}, 16 (2010).
\end{thebibliography}
\end{document}